\documentclass[pra,superscriptaddress,showpacs,preprintnumbers,eqsecnum,amsmath,amssymb]{revtex4}

\begin{document}

\title{Geometrical view of the Mean King Problem}

\author{M. Revzen}
\affiliation {Department of Physics, Technion - Israel Institute of Technology, Haifa
32000, Israel}

\date{\today}

\begin{abstract}
Finite geometry is used to underpin finite, $d^2$, dimensional Hilbert space accommodating
two particles, d dimensional each. d=prime $\ne2$. Central role is allotted to states with
mutual unbiased bases (MUB) labelling underpinned with  points of  finite dual affine plane
geometry (DAPG).  The DAPG lines are shown to underpin maximally entangled states which
form an orthonormal basis spanning the space and provide a novel, geometrical view to a new
solution of the Mean King Problem (MKP). Brief expositions to the topics considered: MUB,
DAPG and the MKP are included rendering the paper self contained.

\end{abstract}

\pacs{03.65.Ta;03.65.Wj;02.10.Ox}

\maketitle

\section {  Introduction}

Several recent studies
\cite{bengtsson,bengtsson1,berge2,vourdas3,saniga,planat1,planat2,combescure,wootters4}
consider the affinity of finite d-dimensional Hilbert space to finite Galois fields, GF(d),
and thereby to finite geometry. These interrelations are of interest as they illuminate
both subjects. The present work contains a novel intuitive geometrical underpinning for the
MUB structure of $d^2$ dimensional Hilbert space accommodating two d-dimensional particles.
(d=prime ($\ne 2$).) The study gives for the first time, to our knowledge, explicit
formulae that relate lines and points of the geometry to {\it states} \cite{rev1},
allowing a geometric view of the relation between product and maximally entangled states \cite{ent}.\\
The analysis underpins Hilbert space states (and operators) with geometrical points and
lines. In particular we show in section III that addition of states in the Hilbert space
may be associated with geometrical requirements leading thereby to the appearance of a
particularly simple balancing term. The uncanny suitability of mutually unbiased bases
(MUB)
 labelling for a convenient coordination scheme is outlined in section IV. In section V we
 give the
 central result of this letter, i.e. the demonstration that the state underpinned with
 geometrical line is a maximally entangled state of remarkable attribute: the expectation
 value of a two particles projector in this state is definite - it is unity (disregarding
  normalization)
 if the underpinning point
 is on the line, nil otherwise. This holds for the two particles projectors while the constituent
 single particle projectors do not commute (being MUB projectors). This attribute leads to
 a novel solution to the Mean King problem outlined in section VI.

\section{   Finite dimensional Mutually Unbiased Bases (MUB) Brief Review}

In a finite, d-dimensional, Hilbert space two complete, orthonormal vectorial bases, ${\cal
B}_1,\;{\cal B}_2$,
 are said to be MUB if and only if (${\cal B}_1\ne {\cal B}_2)$

\begin{equation}
\forall |u\rangle,\;|v \rangle\; \epsilon \;{\cal B}_1,\;{\cal B}_2 \;resp.,\;\;|\langle
u|v\rangle|=1/\sqrt{d}.
\end{equation}
The physical meaning of this is that knowledge that a system is in a particular state in
one basis implies complete ignorance of its state in the other basis.\\
Ivanovich \cite{ivanovich} proved that there can be at most d+1 MUB in a d-dimensional
Hilbert space and gave an explicit formulae for the d+1 bases in the case of d=p (prime
number). Wootters and Fields \cite{wootters2} constructed such d+1 bases for $d=p^m$ with m
a positive integer. Variety of methods for construction of the d+1 bases for $d=p^m$ are
now available
\cite{tal,wootters3,klimov2,vourdas}. Our present study is confined to $d=p\;\ne2$.\\
 We now give explicitly the MUB states in conjunction with the algebraically complete
 operators \cite{schwinger,amir} set:
 $\hat{Z},\hat{X}$.  Thus we label the d distinct states spanning the Hilbert space,
 termed
 the computational basis, by $|n\rangle (n=0,1,..d-1; |n+d\rangle=|n\rangle)$ satisfying
\begin{equation}
\hat{Z}|n\rangle=\omega^{n}|n\rangle;\;\hat{X}|n\rangle=|n+1\rangle,\;\omega=e^{i2\pi/d}.
\end{equation}
The d states in each of the d+1 MUB bases \cite{tal,amir}are the states of computational basis and
\begin{equation} \label{mxel}
|m;b\rangle=\frac{1}{\sqrt
d}\sum_0^{d-1}\omega^{\frac{b}{2}n(n-1)-nm}|n\rangle;\;\;b,m=0,1,..d-1.
\end{equation}
Here the d sets labelled by b are the bases and the m labels the states within a basis. We
have \cite{tal}
\begin{equation}\label{tal1}
\hat{X}\hat{Z}^b|m;b\rangle=\omega^m|m;b\rangle.
\end{equation}
For later reference we shall refer to the computational basis (CB) by b=-1. Thus the above
gives d+1 bases, b=-1,0,1,...d-1 with the total number of states d(d+1) grouped in d+1 sets
each of d states. We have of course,
\begin{equation}\label{mub}
\langle m;b|m';b\rangle=\delta_{m,m'};\;\;|\langle m;b|m';b'\rangle|=\frac{1}{\sqrt d},
\;\;b\ne b'.
\end{equation}
The MUB set is closed under complex conjugation:
\begin{equation}\label{cc}
\langle n|m,b\rangle =\langle \tilde{m},\tilde{b}|n\rangle,
\end{equation}
with $|\tilde{m},\tilde{b}\rangle = |d-m,d-b\rangle$ as can be verified by inspection.
This completes our discussion of MUB.\\

 \section{   Finite Geometry and Hilbert Space Operators}

We now briefly review the essential features of finite geometry required for our study
\cite{bennett,grassl,diniz,shirakova,tomer,wootters4,berger}.\\
A finite plane geometry is a system possessing a finite number of points and lines. There
are two kinds of finite plane geometry: affine and projective. We shall confine ourselves
to affine plane geometry (APG) which is defined as follows. An APG is a non empty set whose
elements are called points. These are grouped in subsets called lines subject to:\\
1. Given any two distinct points there is exactly one line containing both.\\
2. Given a line L and a point S not in L ($S \not\in L$), there exists exactly one line
$L'$ containing S
such that $L \bigcap L'=\varnothing$. This is the parallel postulate.\\
3. There are 3 points that are not collinear.\\
It can be shown \cite{bennett,diniz,shirakova} that for $d=p^m$ (a power of prime) an APG
can be constructed (our study here is for d=p). Furthermore The existence of APG implies
\cite{bennett,diniz,grassl,shirakova}the existence of its dual geometry DAPG wherein the
points and lines are interchanged. Since we shall study extensively this, DAPG, we list its
necessarily built in properties  \cite{bennett,shirakova}. We shall refer to these
by DAPG(y):\\
a. The number of lines is $d^2$, $L_j,\;j=1,2....d^2.$ The number of points is d(d+1),
$S_{\alpha},\;{\alpha = 1,2,...d(d+1)}.$\\
b. A pair of points on a line determine a line uniquely. Two (distinct) lines share one and
only
one point.\\
c. Each point is common to d lines. Each line contain d+1 points.\\
d. The d(d+1) points may be grouped in mutually exclusive sets, d points each with no two
points of a set sharing a line. Such a set is designated by $\alpha' \in \{\alpha \cup
M_{\alpha}\},\; \alpha'=1,2,...d$. ($M_{\alpha}$ contain all the points not on a line with
 $\alpha$ - they are not connected among themselves.) i.e. such a set contain d disjoined
(among themselves) points. These are equivalent classes of the geometry \cite{bennett}.
There are d+1 such sets:
$$\bigcup_{\alpha=1}^{d(d+1)}S_{\alpha}=\bigcup_{\alpha=1}^d R_{\alpha};\;\;
R_{\alpha}=\bigcup_{\alpha'\epsilon\alpha\cup M_{\alpha}}S_{\alpha'};\;\; R_{\alpha}\bigcap
R_{\alpha'}=\varnothing,\;\alpha\ne\alpha'.$$ e. Each point of a set of disjoint points is
connected to every other point not in its set.\\
DAPG(c) allows the definition, which we adopt, and which acquire a meaning upon setting the
points ($S_{\alpha}$) and the lines ($L_j$) as underpinning Hilbert space entities (e.g.
projectors or states, to be specified later) for whom addition is defined:
\begin{equation}\label{A}
S_{\alpha}=\frac{1}{d}\sum_{j\in\alpha}^{d} L_j.
\end{equation}
This implies,
\begin{equation}\label{R}
\sum_{\alpha' \in \alpha \cup M_{\alpha}}^{d}S_{\alpha'}=\frac{1}{d}\sum^{d^2}_{j} L_j,
\end{equation}
and hence require
\begin{equation}
\sum_{\alpha' \in \alpha \cup M_{\alpha}}^{d}S_{\alpha'}\equiv {\cal{R}},
\;\;{\it{independent}}\; of\; \alpha.
\end{equation}

 Eq.(\ref{R}) implies, via DAPG(d)

\begin{equation}\label{I}
{\cal{R}}=\sum_{\alpha' \in \alpha \cup
M_{\alpha}}^{d}S_{\alpha'}=\frac{1}{d}\sum^{d^2}_{j} L_j=
\frac{1}{d+1}\sum^{d(d+1)}_{\alpha}S_{\alpha}.
\end{equation}
This equation, Eq(\ref{I}), reflects relation among equivalent classes within the geometry
\cite{bennett}. It will be referred to as the balance formula: the quantity ${\cal{R}}$
serves as a balancing term, thus, Eqs.(\ref{A}),(\ref{I}) imply,
\begin{equation}\label{line}
L_j=\sum_{\alpha \in j}^{d+1}S_{\alpha} - {\cal{R}}.
\end{equation}

 A particular arrangement of lines and points that satisfies DAPG(x),
x=a,b,c,d,e is referred to as a realization of DAPG. \\
This complete our review of finite geometry.

\section{ Realization of DAPG}

We now consider a particular realization of DAPG of dimensionality $d=p,\ne 2$ which is the
basis of our present study. We arrange the aggregate of the d(d+1) points, $\alpha$, in a
$d\cdot(d+1)$matrix like rectangular array of d rows and d+1 columns. Each column is made
of a set of d points  $R_{\alpha}=\bigcup_{\alpha'\epsilon\alpha\cup
M_{\alpha}}S_{\alpha'};$  DAPG(d). We label the columns by b=$\ddot{0}$,0,1,2,....,d-1 and
the rows by m=0,1,2...d-1.( Note that the first column label of $\ddot{0}$ is for
convenience and does not relate to a numerical value. It designates the computational
basis, CB.) Thus $\alpha=m(b)$ designate a point by its row, m, and its column, b; when b
is allowed to vary . We label the left most column by b=$\ddot{0}$ and with increasing
values of b, that will relate to the basis label, as we move to the right. Thus the right
most column is b=d-1. The top most point in each column is labelled by m=0 with  m values
increasing as one moves to lower rows - the bottom row being m=d-1. The underpinning's
schematics for d=3 is illustrated by the matrix below ( in the matrix below, A stands for
the Hilbert space entity being underpinned with coordinated point, (m,b). In \cite{rev1} A
represented an MUB projector: $A_{\alpha=(m,b)}=\hat{A}_{\alpha}=|m,b\rangle\langle b,m|$.
In the present paper A will be seen to signify a two particles' state to be specified in a
subsequent section).
\[ \left( \begin{array}{ccccc}
m\backslash b&\ddot{0}&0&1&2 \\
0&A_{(0,\ddot{0})}&A_{(0,0)}&A_{(0,1)}&A_{(0,2)}\\
1&A_{(1,\ddot{0})}&A_{(1,0)}&A_{(1,1)}&A_{(1,2)}\\
2&A_{(2,\ddot{0})}&A_{(2,0)}&A_{(2,1)}&A_{(2,2)}\end{array} \right)\].\\

( in the matrix above, A stands for the Hilbert space entity being underpinned with
coordinated point, (m,b). In \cite{rev1} A represented an MUB projector:
$A_{\alpha=(m,b)}\rightarrow\hat{A}_{\alpha}=|m,b\rangle\langle b,m|$. In the present paper
A will be seen to signify a two particles' state to be specified in a subsequent section).

We now assert that the d+1 points, $m_j(b), b=0,1,2,...d-1,$ and $m_j(\ddot{0})$, that form
the line j which contain the two (specific) points $m(\ddot{0})$ and m(0) is given by (we
forfeit the subscript j - it is implicit),
\begin{equation}\label{m(b)}
m(b)=\frac{b}{2}(c-1)+m(0),\;mod[d]\;\;b\ne \ddot{0};\;\;m(\ddot{0})=c/2\;mod[d] .
\end{equation}

The rationale for this particular form is
clarified in the next section. Thus a line j is parameterized fully by
$j=(m(\ddot{0}),m(0))$. We now prove that the set $j=1,2,3...d^2$ lines
covered by Eq.(\ref{m(b)}) with the points as defined above form a realization of DAPG.\\
\noindent 1. Since each of the  parameters, $m(\ddot{0})$ and m(0), can have d values - the
number of lines $d^2$. The number of points in a line is evidently d+1 - one in each
column:   The linearity of the equation precludes having two points with a common value
of b on the same line. DAPG(a).\\
\noindent 2. Consider two points on a given line, $m(b_1),m(b_2);\;b_1\ne b_2$. We have
from Eq.(\ref{m(b)}), ($b\ne \ddot{0},\;b_1 \ne b_2$)
\begin{eqnarray}\label{twopoints}
m(b_1)&=&\frac{b_1}{2}(c-1)+m(0),\;\;mod[d]\nonumber\\
m(b_2)&=&\frac{b_2}{2}(c-1)+m(0),\;\;mod[d].
\end{eqnarray}
These two equations determine uniquely ({\it for d=p, prime}) $m(\ddot{0})$ and m(0). DAPG(b).\\
\noindent For fixed point, m(b), c and m(0) are interrelated, $c\Leftrightarrow m(0),$ thus
the number of free parameters is d (the number of points on a fixed column). Thus each
point is common to d lines. That
the line contain d+1 points is obvious. DAPG(c).\\
\noindent 3. As is argued in 1 above no line contain two points in the same column (i.e.
with equal b). Thus the d points, $\alpha,$ in a column form a set
$R_{\alpha}=\bigcup_{\alpha'\epsilon\alpha\cup M_{\alpha}}S_{\alpha'},$ with trivially
$R_{\alpha}\bigcap R_{\alpha'}=\varnothing,\;\alpha\ne\alpha',$ and
$\bigcup_{\alpha=1}^{d(d+1)}S_{\alpha}=\bigcup_{\alpha=1}^d R_{\alpha}.$ DAPG(d).\\
\noindent 4. Consider two arbitrary points {\it not} in the same set, $R_{\alpha}$ defined
above: $m(b_1),\;m(b_2)\;\;(b_1\ne b_2).$ The argument of 2 above states that, {\it for
d=p}, there is a unique solution for the two parameters that specify the line containing
these points. DAPG(e).\\
We illustrate the above for d=3. The point m(1) is gotten from
$$ m(1)= \frac{1}{2}(2-1)+2=1\;\;mod[3]\;\;\Rightarrow\;m(1)=(1,1).$$
 The full line j labelled by  $j=(\ddot{m},m(0))$ is made up of the 4 points $j:\;1.\big(
m(\ddot{0})=(1,\ddot{0});\;2.m(0)=(2,0);\; 3.m(1)=(1,1)$ and $4.m(2)=(0,2)\big).$ (We shall
denote $m\ddot{0}$ by $\ddot{m}$ when no confusion should arise.) The bracketed numbers
give the point's coordinates.
\[ \left( \begin{array}{ccccc}
m\backslash b&\ddot{0}&0&1&2 \\
0&\cdot&\cdot&\cdot&(0,2)\\
1&(1,\ddot{0})&\cdot&(1,1)&\cdot\\
2&\cdot&(2,0)&\cdot&\cdot\end{array} \right)\].\\
The general formula for a line is:
\begin{equation}\label{l1}
m(b)=m(0)+b/2(2\ddot{m}-1).
\end{equation}
In \cite{rev2} we showed that this  formula for the underpinning line is, equivalently, the
one for equal matrix elements dwelling on a straight line perpendicular to the diagonal, cf.
Appendix B for details.
\begin{equation}
\langle n|\hat{A}_{(m,b)}|n'\rangle=\langle n|\hat{A}_{(m',b')}|n'\rangle,\;b'\ne b;
\end{equation}
with $n+n'=2\ddot{m}$. (The balance formula, Eq.(\ref{I}), in \cite{rev1}, has
${\cal{R}}=\Bbb{I}$.)

\section{  Geometric Underpinning of Two particles' States}

We now consider DAPG underpinning for {\it states} of the Hilbert space of {\it two}
d-dimensional particles. Our coordination scheme is as outlined above
$\alpha=(m,b);\;j=(\ddot{m},m(0)),\; m(b)=m(0)+b/2(2\ddot{m}-1)$ however now each point
will refer to a two particles' {\it state} as specified below. We have thus,

\begin{equation}
|A\rangle_{\alpha};\;\;\alpha= 1,2....d(d+1),\;\;|P\rangle_j;\;\; j=1,2,..d^2.
\end{equation}

$|A_{\alpha}\rangle$ are underpinned  with the d(d+1) points, $S_{\alpha}$ while the
$|P_j\rangle$ with the $d^2$ lines, $L_j$.\\
We define the  states underpinned by the geometrical points , $|A_{\alpha}\rangle$, by

\begin{equation}\label{tilde}
|A_{\alpha}\rangle \equiv |m,b\rangle_1|\tilde{m},\tilde{b}\rangle_2.
\end{equation}
 $|\tilde{m},\tilde{b}\rangle$ is given by Eq.(\ref{cc}).

 With this we return to
states interrelation implied by the geometry: Eqs.(\ref{A}),(\ref{line}) now read
\begin{equation}\label{oprel3}
|A_{\alpha}\rangle=\frac{1}{d}\sum_{j\in
\alpha}^{d}|P_j\rangle\;\;\rightarrow\;\;|P_j\rangle=\sum_{\alpha \in
j}|A_{\alpha}\rangle\; -\;\sum_{\alpha'\in \alpha \cup M_{\alpha}}|A_{\alpha'}\rangle.
\end{equation}

( note that $|P_j\rangle$ is not normalized.) \\

We now show that the choice , $\tilde{m}=d-m,\;\tilde{b}=d-b$ renders, with  the above
underpinning scheme  balance formula, Eq. (\ref{I}) satisfied : Consider,
\cite{fivel1,berge2}, utilizing
\begin{eqnarray}
\sum_{\alpha'\in \alpha \cup M_{\alpha}}|A_{\alpha'}\rangle\equiv\sum_{m\in b}^d|A_{m,b}\rangle&=&\sum_{m\in
b}^d|m,b\rangle_1|\tilde{m},\tilde{b}\rangle_2=\sum_{n}^{d}|n\rangle_1|n\rangle_2\,\nonumber \\
=\sum_{m,n_1,n_2}^d|n_1\rangle|n_2\rangle\langle n_1|m,b\rangle\langle
n_2|\tilde{m},\tilde{b}\rangle_2\;&=& {\cal{R}}\; \;\;independent\; of\;b\; \forall\;b.
\end{eqnarray}
This of course includes the first column, $b=\ddot{0},$ with the "point" in
the $n'$ row underpinning the state $|n'\rangle_1|n'\rangle_2$.\\

The relation among the matrix elements of projectors, $\hat{A}_{(m,b)}=|m,b\rangle\langle
b,m|,$ residing on the line, Eq.(\ref{l1}),\cite{rev1,rev2}, with the two particle states,
$|A_{(m,b)}=|m,b\rangle_1|\tilde{m},\tilde{b}\rangle_2,$ residing on the equivalent line,
Eq.(\ref{l1}), are now used to obtain an explicit formula for
$|P_{j=(\ddot{m},m(0))}\rangle$ ( cf. Appendix B, \cite{rev2}, for further details).

\begin{eqnarray}
|P_{j=\ddot{m},m(0)}\rangle&=&\frac{1}{\sqrt d}\big(\sum_{m(b)\in j}|m,b\rangle_1|\tilde{m},\tilde{b}\rangle_2-|{\cal{R}}\rangle\big)= \nonumber \\
=\frac{1}{\sqrt d}\sum_{n,n'}|n \rangle_1|n'\rangle_2\big[\langle n\sum_{m(b)\in j}\hat{A}_{m,b}\;-\;\Bbb{I}|n'\rangle\big]&=&\frac{1}{\sqrt d}\sum_{n,n'}|n \rangle_1|n'\rangle_2\delta_{n+n',2\ddot{m}}\omega^{-(n-n')m(0)}\;\;\forall\;b,
\end{eqnarray}
where we used the results of appendix B. The expression for the line state will be put now in a more pliable form for our analysis,
\begin{eqnarray}\label{fivel}
|P_{j=\ddot{m},m(0)}\rangle&=&\frac{1}{\sqrt d}\sum_{n,n'}|n\rangle_1|n'\rangle_2\delta_{n+n',2\ddot{m}}\omega^{-(n-n')m(0)}= \nonumber \\
=\frac{\omega^{2\ddot{m}m(0)}}{\sqrt d}\sum_{n}|n\rangle_1|2\ddot{m}-n\rangle_2\omega^{-2nm(0)}&=&\frac{\omega^{2\ddot{m}m(0)}}{\sqrt d}\sum_{n}|n\rangle_1\hat{X}^{2\ddot{m}}\hat{Z}^{2m(0)}{\cal{I}}|n\rangle_2 = \nonumber \\
&=&\frac{\omega^{2\ddot{m}m(0)}}{\sqrt d}\sum_{m}|m,b\rangle_1\hat{X}^{2\ddot{m}}\hat{Z}^{2m(0)}{\cal{I}}
|\tilde{m},\tilde{b}\rangle_2.
\end{eqnarray}
The inversion operator ${\cal{I}}$ is defined via
${\cal{I}}|n\rangle=|-n\rangle=|d-n\rangle$. $\hat{X},\hat{Z}$ were defined in section II.
The orthonormality of $|P_j\rangle$ is proved in appendix A.

The central result of our geometrical underpinning is the following
\begin{eqnarray}
\langle m,b|_1\langle\tilde{m},\tilde{b}|_2P_{j=\ddot{m},m(0)}\rangle&=&
\frac{\omega^{(2b\ddot{m}^2-b\ddot{m})}}{\sqrt
d}\delta_{m,\big(m(0)+b/2[2\ddot{m}-1]\big)},\;\;b\ne \ddot {0},\nonumber \\
\langle n|_1\langle n|_2 P_{j=\ddot{m},m(0)}\rangle&=&\frac{1}{\sqrt 2}\delta_{n,\ddot{m}},\;\;b=\ddot {0}, \;i.e.\;computational\;basis.
\end{eqnarray}
Thus, if  $ m\ne m(0)+b/2(2\ddot{m}-1)$, it vanishes. i.e. only if the point (m,b) is on
the line j the overlap is non zero. This is a remarkable attribute: Each and every one of
the observables $|m,b\rangle_1|\tilde{m},\tilde{b}\rangle_2\langle
\tilde{b},\tilde{m}|_2\langle b,m|_1$ has a {\it definite and known} value if measured in
the state $|P_{(\ddot{m},m(0))}\rangle$ yet these observables do {\it not} commute. The
value obtained is 1/d if the point is on the line, nil otherwise. This allows a new
approach to the Mean King problem which we now outline.

\section{The Mean King Problem}

The Mean King Problem (MKP), initiated by \cite{lev1,bg1},  was analyzed in several
publications (see the comprehensive list given in \cite{berge2}). Briefly summarized it
runs as follows. Alice may prepare a state to her liking. The King measures its MUB state.
He does not inform Alice of his observational results nor of the basis he used. At this
juncture Alice perform a control measurement of her choice so as to accommodate the
following requirement:  {\it{After}} completing her control measurement she  will be told
by the King the {\it{basis}}, b,  that he used in his measurement.  She must now  deduce
with no further measurements the actual state (m,b) that he observed.\\

The novel solution we provide  has an intuitive geometrical meaning which we give as
follows. The state that Alice prepare is one of the line vectors,
$|P_{j=(\ddot{m},m(0))}\rangle$. Thus she knows both $\ddot{m}$ and m(0). The King's
measurement is along a line of some fixed b so it picks a point on the line above. Formally
the Kings measurement along some b (basis) yielding m projects the state $|P_j\rangle$ to
$|m,b\rangle\hat{X}^{2\ddot{m}}\hat{Z}^{2m(0)}{\cal{I}}|\tilde{m},\tilde{b}\rangle$. Now
Alice measures the lines that intersect this point - recalling, see Appendix A, that the
$d^2$ vectors $|P_j\rangle$ are orthonormal, she measures the {\it non}degenerate operator,
$$\sum_{j}^{d^2}|P_j\rangle \gamma_j\langle P_j|$$
to get say $j'=(\ddot{m}',m'(0)).$ Thus, using Eq.(\ref{fivel})
\begin{eqnarray}
\langle P_{j'=(\ddot{m}',m'(0))}|m,b\rangle\hat{X}^{2\ddot{m}}\hat{Z}^{2m(0)}{\cal{I}}
|\tilde{m},\tilde{b}\rangle\;
\ne 0&\rightarrow& \nonumber \\
\langle \tilde{m},\tilde{b}|{\cal{I}}\hat{X}^{-2\ddot{m}'}\hat{Z}^{-2m'(0)}\hat{X}^{2\ddot{m}}
\hat{Z}^{2m(0)}{\cal{I}}
|\tilde{m},\tilde{b}\rangle\;\ne 0&\rightarrow& \nonumber \\
m=m(0)-m'(0)+\frac{b}{2}(\ddot{m}-\ddot{m}').
\end{eqnarray}
Hence knowing $m(0),m'(0),\ddot{m},\ddot{m}'$ upon being informed of b Alice can calculate m
appeasing the King.

\section {   Summary and Concluding Remarks}
We outlined finite plane geometrical underpinning to {\it states} of a $d^2$ dimensional
Hilbert space accommodating two particles each spanning a d dimensional space, for d=prime
($\ne 2$). The geometrical points were coordinated with (i.e. assigned values of) mutual
unbiased bases (MUB) states labels: $|A_{m,b}\rangle$, m denotes the vector in the base b,
it locates the vertical coordinate within the column labeled by b giving the position of
the column. Points of the geometry $\alpha=(m,b)$ underpin product states
$|A_{\alpha}\rangle=|m,b\rangle_1|\tilde{m},\tilde{b}\rangle_2,$ 1 and 2 refers to the
particles and $\tilde{m}=d-m,\;\tilde{b}=d-b.$ The state vector
$|P_{j=(\ddot{m},m(0))}\rangle$ is underpinned by a line $L_j$ of the geometry and is
conveniently labelled by two points that reside on the line: $\ddot{m}$ denoting the point
on the computational basis column and m(0) the point in the column of
eigenfunctions of the displacement operator.\\
The states $|P_j\rangle$ turn out to be maximally entangled states in possession of remarkable
attribute: expectation values of the projectors $|A_{\alpha}\rangle\langle A_{\alpha}|$ in these
states gives the coordinate of the point $\alpha$ on the line j, if the point on it, it vanish otherwise.
 Thus the state $|P_j\rangle$ yields definite values to two particles projectors with its constituent
 single particle projectors {\it non} commuting. This was used to outline a novel, geometrically based,
solution to the Mean king Problem.

\section*[Appendix A] {Appendix A: Orthogonality of $|P_j\rangle$}

Noting that $\langle{\cal{R}}|{\cal{R}}\rangle=d$ and $\langle P_j|{\cal{R}}\rangle=d+1$,
We get, for j=j':

\begin{equation}
\langle P_j|P_j\rangle= \frac{1}{d}\{\sum_{\alpha \in j}\langle
A_{\alpha}|-\langle{\cal{R}|\}}\{\sum_{\alpha' \in
j}|A_{\alpha'}\rangle-|{\cal{R}}\rangle\}=\frac{1}{d}\{1+d +\sum_{\alpha\ne
\alpha'}^{d+1}[\langle A_{\alpha}|A_{\alpha'}\rangle]-2(d+1)+d\}=1
\end{equation}
Where we used that $\sum_{\alpha\ne\alpha'}^{d+1}[\langle
A_{\alpha}|A_{\alpha'}\rangle=(d+1)\frac{d}{d}=d+1$.

For $j\ne j'$ the geometry dictates, DAPG(b), that distinct lines share {\it one} point.
Thus the first term above is 1 rather than 1+d hence
$$\langle P_j|P_j \rangle=0, \;\;j \ne j'$$
i.e.
\begin{equation}
\langle P_j|P_j'\rangle=\delta_{j,j'}.\;\;QED
\end{equation}

\section*[Appendix A] {Appendix B: The Line Equation}

Let $\hat{A}_{\alpha}=|m,b\rangle\langle b,m|;\;\alpha=(m,b).$ Eq.(\ref{mxel}) gives
\begin{equation}
\langle n|\hat{A}_{\alpha}|n'\rangle=\frac{1}{d}\omega^{(n-n')[b/2(n+n'-1)-m_b]}.
\end{equation}
Requiring $\langle n|\hat{A}_{\alpha}|n'\rangle=\langle n|\hat{A}_{\alpha"}|n'\rangle,\;n\ne n'$ implies
\begin{equation}
b/2(n+n'-1)-m_b=b"/2(n+n'-1)-m"_{b"}\rightarrow \;m(b)=m(0)+\frac{b}{2}(2\ddot{m}-1),
\end{equation}
upon setting b"=0 and choosing $n+n'=\ddot{m}$. This gives the line equation, i.e. it gives the values of the row m in the column b of the line  $j=(\ddot{m},m(0))$. Matrix elements wherein $n+n'\ne 2\ddot{m}$
are unequal thus upon summing over $\alpha \in j$ they add up to nil as they are the d roots of unity.
\begin{equation}
\langle n|\sum_{\alpha\in j}\hat{A}_{\alpha}\;-\;\Bbb{I}|n'\rangle=\delta_{n+n',2\ddot{m}}\omega^{(n-n')m(0)}.
\end{equation}
As the sole nonzero diagonal element is unity at $n=n'=\ddot{m}$ - all other diagonal
element being removed by the  subtraction of the balancing term - the unit matrix,
$\Bbb{I}$.

Acknowledgments: Informative discussions with Prof. A. Mann and P.A. Mello are gratefully acknowledged.\\


\begin{thebibliography}{999}




\bibitem{wootters4} W. K. Wootters, Found. of Phys. {\bf 36}, 112 (2006).
\bibitem{vourdas} A. Vourdas, Rep. Math. Phys. {\bf 40}, 367 (1997), Rep. Prog. Phys. {\bf 67}, 267 (2004).
\bibitem{vourdas3} A. Vourdas, J. Phys. A {\bf 40} R285 (2007).
\bibitem{saniga} M. Saniga, M. Planat and H. Rosu, J. Opt. B: Quantum Semiclassic Opt. {\bf
6} L19 (2004).
\bibitem{planat1} M. Planat and H. C. Rosu, Europ. Phys. J. {\bf 36}, 133 (2005).
\bibitem{planat2} M. Planat, H. C. Rosu and S. Perrine, Foundations of Physics {\bf 36}, 1662 (2006).
\bibitem{combescure} M. Combescure, quant-ph/0605090 (2006).

\bibitem{wootters2} W. K. Wootters and  B. D. Fields, Ann. Phys. (N.Y.) {\bf 191}, 363 (1989).
\bibitem{wootters3} K. S. Gibbons, M. J. Hoffman and W. K. Wootters, Phys. Rev. A {\bf 70}, 062101 (2004).

\bibitem{bennett} M. K. Bennett {\it Affine and Projective geometry} John Wiley , sons, Inc. (1995).
\bibitem{tomer} T. Bar-On, Jour. Math. Phys. {\bf 50}, 072106 (2009).
\bibitem{berge2} T. Durt, B.-G. Englert, I Bengtsson and K. Zyczkowski, International J. of Quantum Information, {\bf 8}, 535 (2010).
\bibitem{lev1} L. Vaidman, Y. Aharonov and D. Albert, Phys.Rev.Let. {\bf 58}, 1385 (1987).
\bibitem{bg1} B,-G. Englert and Y. Aharonov, Phys. Lett. A {\bf 284}, 1  (2001).
\bibitem{amir} A. Kalev, M. Revzen and F. C. Khanna, Phys. Rev. A {\bf 80}, 022112 (2009).
\bibitem{berger} M. Berger, {\it Geometry Revealed}. Springer-Verlag, Berlin (2010)
\bibitem{bengtsson} I. Bengtsson, quant-ph/0406174v2, (2004) .
\bibitem{bengtsson1} I. Bengtsson and K. Zyczkowski {\it Geometry of Quantum States}
Cambridge University press, Cambridge, (2006).
\bibitem{rev1} M. Revzen, EPL, {\bf 98}, 10001 (2012).
\bibitem{rev2} M. Revzen, quant-ph/1111.6446v4 (2011).

\bibitem{fivel1} D. I. Fivel {\it Fundamental Problems in Quantum Mechanics} Ed. D.M.
Greenberger, A. Zeilinger, Ann. NY Academy of Science {\bf 755}, (1995). D. I. Fivel Phys.
Rev. Lett. {\bf 74}, 835 (1995).










\bibitem{ent} M. Revzen, Phys. Rev. A {\bf 81}, 012113 (2010).

\bibitem{schwinger} J. Schwinger, Proc. Nat. Acad. Sci. USA {\bf 46}, 560 (1960).










publication Inc.



\bibitem{ivanovich} I. D. Ivanovic, J. Phys. A, {\bf 14}, 3241 (1981).
\bibitem{tal} S. Bandyopadhyay, P. O. Boykin, V. Roychowdhury and F. Vatan,
Algorithmica {\bf 34}, 512 (2002).




\bibitem{klimov2} A. B. Klimov, C. Munos and J. L. Romero, quant-ph/0605113 (2005).
{\bf 45}, 2171 (2004).
\bibitem{grassl} M. Grassl, Electronic Notes in Discrete Mathematics {\bf 20}, 151 (2005).





\bibitem{diniz} F. J. MacWilliams and N. J. A. Sloane, {\it The Theory of Error Correcting
Codes}, North Holland, Amsterdam, (1977).
\bibitem{shirakova} S. A. Shirakova, Russ. Math. Surv. {\bf 23} 47 (1968).









\end{thebibliography}
\end{document}